\documentclass[12pt]{iopart}
\renewcommand\a{\alpha}
\renewcommand\b{\beta}

\newcommand\m{\mu}
\newcommand\n{\nu}

\newcommand{\ovl}[1]{\overline{#1}}

\newcommand{\non}{\nonumber\\}

\newcommand{\bee}{\begin{equation}}
\newcommand{\eee}{\end{equation}}
\newcommand{\bea}{\begin{eqnarray}}
\newcommand{\eea}{\end{eqnarray}}
\newcommand{\ba}[1]{\begin{array}{#1}}
\newcommand{\ea}{\end{array}}

\newcommand{\eqrf}[1]{Eq.(\ref{#1})}

\begin{document}

\title{Kinetic equation for gluons at the early stage}

\author{Qun Wang\dag , Krzysztof Redlich\ddag , 
Horst St\"ocker\dag \; and Walter Greiner\dag }  

\address{\dag\ Institute f\"ur Theoretische Physik, J. W.
Goethe-Universit\"at, D-60054 Frankfurt, Germany }

\address{\ddag\ Theory Division, CERN, 
CH-1211 Geneva 23, Switzerland}

\begin{abstract}

We derive the kinetic equation for pure gluon QCD plasma in a 
general way, applying the background field method. 
We show that the quantum kinetic equation contains a term
as in the classical case, that describes a color charge precession
of partons moving in the gauge field. We emphasize that this new
term is necessary for the gauge covariance of the resulting
equation.

\end{abstract}



%\maketitle

Heavy ion collisions at ultra-relativistic energies are widely
expected to be a laboratory to study the formation and properties
of  highly excited  QCD  matter in the deconfined Quark-Gluon
Plasma (QGP) phase \cite{rev1}. The QGP is considered as a
partonic system being at (or close to) local thermal equilibrium.
Thus, to study the conditions for the possible formation of QGP in
heavy ion collisions one needs to address the question of
thermalization of the initially produced partonic medium
\cite{baier}. On the theoretical level this requires the
formulation of kinetic equations \cite{jp} involving color degrees
of freedom and the non-Abelian structure of QCD dynamics.
Different models for the initial conditions in ultra-relativistic
heavy ion collisions suggest that at the early stage the medium is
dominated by gluon degrees of freedom \cite{sat}; the kinetic
equation for the pure gluon plasma is thus of special interest.

The kinetics of a particle moving in a non-Abelian gauge field was
first studied for SU(2) symmetry \cite{wong}. The equation of
motion derived in \cite{wong} made it clear that in the presence
of the gauge field the color charge of the particle keeps rotating
or undergoing precession with time. Consequently, there is an
explicit term in the classical kinetic equation, which describes
the time evolution of the color charge of partons. Applying a
typical method to derive a semi-classical transport equation from
the quantum theory, which is contained in the formulation of the
equation of motion for the Wigner function and its subsequent
gradient expansion \cite{heinz,elze86,elze88}, one loses the
explicit presence of the color charge precession because the 
obtained transport equations are not most simplified. 
The author of Ref.\ \cite{mrow} simplified the kinetic 
equation for gluons based on Ref.\ \cite{elze88}. However  
the simplification is under the assumption of a 
specific form of the covariant Green function (actually it is 
a special case of this work), and the color precession term 
was not explictly and transparently interpreted.

The usual treatment of the gluon transport equation is based on
the decomposition of the gluon field into a mean field and a
quantum fluctuation, or into a soft and a hard part. Under this
approximation the gluon transport equation then describes the
kinetics of the quanta in the classical mean field
\cite{heinz,geiger}. This picture is somehow similar to the one
used while studying the energy loss of a fast parton 
moving in the soft mean field
\cite{mgxw,baier1}. Since this decomposition is not rigorous, the
explicit gauge invariance of the formalism remains unclear. To
include the classical chromofield into QCD in a proper way, one
uses  the background field method of QCD (BG-QCD) introduced by
DeWitt and 't Hooft \cite{dewitt,thooft,abb81}. The advantage of
BG-QCD is that it is formulated in an explicit gauge invariant
manner.

One of the first attempts to derive the gluon transport equation
in BG-QCD was presented in \cite{elze90}. But the obtained
equation is not transparent. The most recent work has
been done by Blaizot and Iancu \cite{blaizot,blaizot1} in the
context of the close-time-path (CTP) formalism. There, however,
the authors focus on formulating the transport equation in the
vicinity of equilibrium.

In this work, we use the CTP method to derive the kinetic
equation for the gluon plasma in BG-QCD. Our derivation is going
beyond previous results as it is quite general and not limited to
the vicinity to equilibrium. One of the most important features of
the kinetic equation derived here
is that a color charge precession term is explicitly present,
similar to the one in the classical equation
\cite{wong,heinz,elze86}. We demonstrate that this term is
necessary to guarantee the gauge covariance of the resulting
Vlasov equation.

In the following we use $g_{\m\n}={\rm diag}(1,-1,-1,-1)$ as the
metric tensor, and we write Lorentz indices as subscripts and
color ones as superscripts for the relevant quantities.
For the gauge  field  and its field tensor we also denote:
$A_{\m}\equiv A_{\m}^aT^a$ and $F_{\m\n}[A]\equiv
F_{\m\n}^a[A]T^a$, where $(T^a)^{ij}=if^{iaj}$ are the generators
of the $SU(3)_c$ adjoint representation.
The two-point Green function (GF) or self-energy (SE) are treated
as matrices, and thus the color  and/or Lorentz indices are
sometimes omitted.

%%%%%% background QCD
Applying the background field method, we decompose the
conventional gluon field into the sum of a classical background
part $A$ and a quantum fluctuation $Q$. Including the appropriate
gauge fixing and ghost term for the background gauge
$D_{\mu}^{ij}[A]Q_{\mu}^j=0$, the BG-QCD Lagrangian reads
\cite{abb81}
%%%%%%%%%%%%%%%%%%
\bea {\cal L}&=&
-\frac 14F_{\mu \nu}^i[A+Q]F_{\mu \nu}^i[A+Q]
-\frac {1}{2\alpha} (D_{\mu}^{ij}[A]Q_{\mu}^j)^2 \non
&&+\overline{C}^i D_{\mu}^{ij}[A]D_{\mu}^{jk}[A+Q]C^k \;,
\label{s} \eea
%%%%%%%%%%%%%%%%%%
where $D_{\mu}^{ij}[A(x)]= [\partial _{x\mu}-igA_{\mu}(x)]^{ij}$
is the covariant derivative, $C^i$/$\overline{C}^i$ are the
ghost/anti ghost fields and $\alpha$ is the gauge fixing
parameter.

The above Lagrangian is invariant under the local gauge
transformation of type I (type II transformation is irrelevant to the 
current problem) where the background field transforms as
a conventional gauge fields, $A'_{\mu}=UA_{\mu}U^{-1} +ig^{-1}
U\partial _{\mu}U^{-1}$, while the gluon and ghost field transform
as a matter field, $Q'_{\mu}=UQ_{\mu}U^{-1}$ \cite{abb81}. Here
$U(x)=\exp [ig\omega ^a(x)T^a]$ is the transformation matrix.

The non-equilibrium dynamics is usually described in the CTP
formalism \cite{ctp}. Here the action can be written as
$S_{CTP}=S_+-S_-+K(A_{\pm},Q_{\pm})$ where all fields in $S_{\pm}$
are defined on the positive/negative time branches and
$K(A_{\pm},Q_{\pm})$ is the kernel incorporating initial state
correlations. The GF for the gluon has four components: $ [
G^{++}, G^{+-}, G^{-+}, G^{--} ]\equiv [ G^{F}, G^{<}, G^{>},
G^{\ovl{F}} ]$, marked by the positive or the negative time,
respectively. At the tree level there are only vertices with all
positive or all negative time variables, which differ by a sign.
The non-local kernel $K(A_{\pm},Q_{\pm})$ appearing in $S_{CTP}$
is non-vanishing only at the initial time and plays the role of
the initial density matrix. If the initial time is not in the
remote past, it introduces a time dependence to the distribution
function, and consequently the time-translational invariance is
lost \cite{qwang}. It is interesting to note that the situation
here is similar to the pinch singularity discussed in
\cite{pinch}.

In the CTP formalism, choosing the physical representation for the
GF and SE \footnote{ In the physical representation the elements
of GF and SE matrices are expressed by the symmetric (C), the
retarded (R) and the advanced (A) components. They are related to
the original components by the following transformations:
$G^A=G^F-G^>$, $G^R=G^F-G^<$, $G^C=G^>+G^<$ (the same relations
hold for the SE $\Pi ^{C/R/A}$ up to a sign in front of $\Pi ^>$
and $\Pi ^<$).}, the Dyson-Schwinger equation (DSE) for $G^C$
reads
%%%%%%%%%%%%%%%%%%%%%
\bea {\mathcal D}(x_1)G^C(x_1,x_2) &=&-\int d^4x'[\Pi
^C(x_1,x')G^A(x',x_2) \non &&+\Pi ^{R}(x_1,x')G^{C}(x',x_2)]\;,
\label{sch-dy-c1}\\
G^C(x_1,x_2){\mathcal D}^{\dagger}(x_2) &=&-\int
d^4x'[G^R(x_1,x')\Pi ^C(x',x_2) \non &&+G^{C}(x_1,x')\Pi
^{A}(x',x_2)]\;,
\label{sch-dy-c2} \eea
%%%%%%%%%%%%%%%%%%%%%
where the differential operators ${\cal D}$ and ${\cal
D}^{\dagger}$ in the Feynman gauge \footnote{ In fact due to  gauge
conditions: $D^{ij}_{\mu}[A(x_1)]G_{\mu\nu}^{jk}(x_1,x_2)=0$ and
$G^{ij}_{\mu\nu}(x_1,x_2)D_{\nu}^{\dagger ;jk}[A(x_2)]=0$, Eqs.
(\ref{sch-dy-c1},\ref{sch-dy-c2}) are independent of the gauge
parameter $\a $.} ($\alpha =1)$ are expressed by:
%defined by
%%%%%%%%%%%%%%%%%%%%%%%%%%%%%
 \bea
{\mathcal D}^{hi}_{\rho\sigma}
%&=&g_{\rho\sigma}D_{\mu}^{ha}[A]D_{\mu}^{ai}[A]
%-D_{\sigma}^{ha}[A]D_{\rho}^{ai}[A]\nonumber\\
%&&+\frac 1{\alpha}D_{\rho}^{ha}[A]D_{\sigma}^{ai}[A]
%+gf^{hai}F^a_{\rho\sigma}[A]\nonumber\\
%&=&g_{\rho\sigma}D_{\mu}^{ha}[A]D_{\mu}^{ai}[A]
&=&g_{\rho\sigma}D_{\mu}^{ha}[A]D_{\mu}^{ai}[A]
+2gf^{hai}F^a_{\rho\sigma}[A]\;,
\label{d1}\\
{\mathcal D}^{\dagger;hi}_{\rho\sigma}
&=&g_{\rho\sigma}D_{\mu}^{\dagger;ha}[A]D_{\mu}^{\dagger;ai}[A]
+2gf^{hai}F^a_{\rho\sigma}[A] \;,
\label{d2} \eea
%%%%%%%%%%%%%%%%
where $D_{\mu}^{\dagger;ij}[A(x)] \equiv
[ \stackrel{\leftarrow}{\partial }_{x\mu}
+igA_{\mu}(x) ]^{ij}$
is the conjugate covariant derivative
where the differential operator acts
on the function on its left.

In the evolution of the gluon field one distinguishes different
scales, which characterize quantum and soft collective motion. We
introduce a mass parameter, $\m$, as the separation point of the
quantum and the kinetic scale. In the weak coupling limit $g\ll
1$, the scale of collectivity $\sim 1/(g\mu)$ is much larger than
the typical extension of hard quantum fluctuations $\sim 1/\mu$.
The effect of the classical field $A$ on the hard quanta involves
the coupling $gA$ to the hard propagator and is of the size of the
soft wavelength $\sim 1/(g\mu)$. The above separation of scales is
the basis for the gradient expansion where one expresses all
2-point GF$^,$s in terms of the relative $y$ and the central $X$
coordinate. Here are some typical scales: $y=x_1-x_2\sim 1/\mu$,
$X=(x_1+x_2)/2\sim 1/(g\mu)$, $A(X)\sim \mu$ and $F[A(X)]\sim
g\mu^2$.

In order to obtain the gauge-covariant
kinetic equation one uses the gauge-covariant Wigner function
$\tilde{G}(X,y)$ ($\tilde{G}^C$, $\tilde{G}^>$ or $\tilde{G}^<$)
defined by
%%%%%%%%%%
\bea
\label{tg-g}
G(x_1,x_2)&=&V(x_1,X)\tilde{G}(X,y)V(X,x_2)\;,
\eea
%%%%%%%%%%
where $V(z_1,z_2)=T_P\exp (ig\int _{P;z_2}^{z_1} dz_{\mu}A_{\mu})$
denotes a Wilson link with respect to
the classical background field.
One can also define the Wilson
link as a functional of $A+Q$,
but it is a much more complicated case and
beyond the scope and subject of this work.

The covariant Wigner function $\tilde{G}(X,y)$ transforms as
$U(X)\tilde{G}(X,y)U^{-1}(X)$;  only the transformation at a
single point $X$ is thus relevant. For $G(x_1,x_2)$, however,  the
gauge transformation involves two points and therefore  is not
gauge-covariant.

Taking the difference between \eqrf{sch-dy-c1} and
\eqrf{sch-dy-c2}, we derive the following 
kinetic equation in terms of the gauge covariant 
Wigner function $\tilde{G}^C(X,y)$:
%%%%%%%%%%%%%%%
\bea
&& q\cdot \partial _X\tilde{G}_{\alpha\gamma}^{C}
+ig(\tilde{G}_{\alpha\gamma}^{C}q\cdot A -q\cdot
A\tilde{G}_{\alpha\gamma}^{C}) \non
&&+\frac 12 gq_{\nu}F_{\nu\lambda}(\partial _{q\lambda}
\tilde{G}_{\alpha\gamma}^{C})
+\frac 12 g(\partial _{q\lambda}\tilde{G}_{\alpha\gamma}^{C})
q_{\nu}F_{\nu\lambda}\non
&&+g(F_{\alpha\beta}\tilde{G}_{\beta\gamma}^{C}
-\tilde{G}_{\alpha\beta}^{C}F_{\beta\gamma})=0\;,
\label{treq1}
\eea
%%%%%%%%%%%%%%%
where  $\tilde{G}_{\alpha\gamma}^{C}\equiv
\tilde{G}_{\alpha\gamma}^{C}(X,q)$ is the Fourier transform of
$\tilde{G}(X,y)$ with respect to $y$:
$\tilde{G}(X,q)=\int
d^4y\,\tilde{G}(X,y)e^{iqy}$.

The above equation is located at the collective coordinate $X$ and
is {\it gauge-covariant} under the local gauge transformation
$U(X)$, i.e. it transforms as $U(\cdots )U^{-1}$. Indeed, noting
that both $F_{\m\n}$ and $\tilde{G}_{\alpha\gamma}^{C}$ are 
gauge-covariant and $\partial_{q\m}$ does not affect $U$, 
it is obvious that the last three terms are gauge-covariant. 
With some algebra, we can also prove the 
first two terms preserve gauge covariance too.

%\section{Color charge precession in kinetic equation}

The quantum kinetic equation \eqrf{treq1} is derived in a quite
general manner without requirements or assumptions that the QCD
plasma appears close to equilibrium. In the following
we show that \eqrf{treq1} is a natural
quantum generalization of the classical Vlasov equation.
In particular the color charge precession
will be explicitly identified in the quantum
description of color charge kinetics given by \eqrf{treq1}.

The classical kinetic equation for the color singlet distribution
function $f(x,p,Q)$ has the following form \cite{wong,heinz,elze86}:
\bea
\label{wongeq} &&p_{\mu}[\partial _{\mu}-gQ^aF_{\mu\nu}^a
\partial _{p\nu} \non
&&-gf^{abc}A_{\mu}^b(x)Q^c\partial _{Q^a}] f(x,p,Q)=0\;,
\eea
%%%%%%%%%%
where $Q^a$ is the classical color charge and  $a=1,.., N^2_c-1$.

Comparing \eqrf{wongeq} with the quantum expression \eqrf{treq1}
it is clear that the color singlet distribution function $f$ is
replaced by the gauge-covariant Wigner function $\tilde{G}^C$
which is a color matrix in the adjoint representation. One can
also recognize that the first, third and fourth terms of
\eqrf{treq1} are the quantum generalization of the first two terms
in \eqrf{wongeq}. The last term in \eqrf{treq1} appears from the
covariant operators \footnote{This term can be written in
different form by using generators of the Lorentz transformation
in {\em vector} representation. Similar term can be found for the
quark, but expressed through generators in {\em spinor}
representation.} and hence  is not present in the classical
equation.

Particularly interesting is the appearance of the second term in
\eqrf{treq1}. We have seen that its presence is crucial to assure
the gauge covariance of the Vlasov equation. In addition this
term has an interesting physical meaning. It is the quantum
analogue to the color charge precession in the classical kinetic
equation. To see this more clearly, one can expand
$\tilde{G}^C_{\a\b}(X,q)$ with respect to the generators $T^a$ of
the adjoint representation as follows \bea
\tilde{G}^C_{\a\b}(X,q)&=&N_{\a\b}(X,q)+T^aN^a_{\a\b}(X,q) \non
&&+T^aT^bN^{ab}_{\a\b}(X,q)+\cdots \;, \label{expansion} \eea
where $T^a$ are quantum analogues to the classical color charges
$Q^a$; $N_{\a\b}(X,q)$ is the color singlet and $N^a_{\a\b}(X,q)$
the color octet function, etc. This expansion is similar to the
multipole expansion of an electromagnetic source in
electrodynamics. In the lowest order approximation, one can keep
only the first two terms in \eqrf{expansion} and neglect all
others. Then the second term of \eqrf{treq1} becomes
\begin{equation}
ig(\tilde{G}_{\alpha\gamma}^{C}q\cdot A -q\cdot
A\tilde{G}_{\alpha\gamma}^{C})\simeq -g f^{abc} q_{\m} A^b_{\m}
T^c \partial_{T^a} \tilde{G}_{\alpha\gamma}^{C}
\end{equation}
which is just the classical
color precession term in Eq.(\ref{wongeq}).

Since we know that $D_X\sim gA(X)\sim g\m$, the first two terms of
\eqrf{treq1}, i.e. $q\cdot \partial
_X\tilde{G}_{\alpha\gamma}^{C}$ and the color precession term, are
at leading order $O(g\m ^2)$, while other terms are at subleading
order $O(g^2\m ^2)$. In the vicinity of equilibrium the natural
scale in the system is the temperature $T$. The mean distance
between particles is of the order of $\sim 1/T$, while $1/(gT)$
characterizes the scale of collective excitations
\cite{blaizot,blaizot1}. For small coupling constant $g$ these two
scales are well separated. The covariant Wigner functions can be
expanded around their equilibrium values:
$\tilde{G}=\tilde{G}^{(0)}+\delta\tilde{G}$, where the equilibrium
function $\tilde{G}^{(0)}$ is a color singlet and the fluctuation
$\delta\tilde{G}\sim g^2\tilde{G}$. Typical scales are $q\sim T$,
$D_X\sim g^2T$, $gF\sim (D_X)^2\sim g^4T^2$. Thus, at leading
order, only the first term of \eqrf{treq1} survives and the
precession term vanishes due to the color-singlet nature of
$\tilde{G}^{(0)}$.

%\section{Summary and conclusions}
%\vskip 0.1cm

In summary, by applying the closed-time-path formalism we have
derived the gluon kinetic equation in the background gauge of QCD.
Equation (\ref{treq1}) is quite general and is not limited  to the
vicinity of equilibrium. It is a kinetic equation with respect to
a gauge-covariant Wigner function, which is a matrix in adjoint
color space. Therefore it contains many non-Abelian features which
are absent from  the well known classical equation. A notable
feature is that,  as in the classical case, it contains a term
that corresponds to the color precession, the non-Abelian analogue
to the Larmor precession for particles with 
magnetic moments in a magnetic field. 
This term was not explicitly shown before
in the formulation of a Boltzmann equation for the QCD plasma. We
find that this term is necessary to the gauge covariance of the
kinetic equation.

\ack
One of us Q.W. acknowledges financial support of the Alexander von
Humboldt-Foundation (AvH) and appreciated help from D. Rischke.
This work is partially funded by DFG, BMBF and GSI. K.R.
acknowledges a partial support of the Polish Committee for
Scientific Research (KBN-2P03B 03018). Stimulating comments and
discussions with R. Baier, J.-P. Blaizot, E. Iancu and S. Leupold
are acknowledged. Our special thanks go to X.-N. Wang for his help
and interest in this work.

%%%%%%%%%%%%%%%%%%%%%%%%%%%%%%%%%%%%%%%%%%%%%%
%%%%%%%%%%%%%%%%%%%%%%%%%%%%%%%%%%%%%%%%%%%
%%%%%%%%%%%%%%%%%    DOCUMENT ENDED %%%%%%%%%%%%%%%%%%%%%%%%%%%%%
%%%%%%%%%%%%%%%%%%%%%%%%%%%%%%%%%%%%%%%%%%%%%
%%%%%%%%%%%%%%%%%    DOCUMENT ENDED %%%%%%%%%%%%%%%%%%%%%%%%%%%%%
%%%%%%%%%%%%%%%%%%%%%%%%%%%%%%%%%%%%%%%%%%%%%


\section*{References}

\begin{thebibliography}{99}

\bibitem{rev1} For a recent review, see eg.
S.A. Bass, M. Gyulassy, H. Stocker and W. Greiner; J. Phys. G{\bf
25}, R1 (1999); H. Satz, Rep. Prog. Phys. {\bf 63}, 1511 (2000).

\bibitem{baier} %This quastion was most recently addresed in:
R. Baier, A.H. Mueller, D. Schiff and D.T. Son, Phys. Lett.
B{\bf 502}, 51 (2001).

\bibitem{jp} For a recent review, see eg.
J.-P. Blaizot and  E. Iancu, hep-ph/0101103, Phys. Rept., to
appear.

\bibitem{sat} L. McLerran and R. Venugopalan,
Phys. Rev. D{\bf 49}, 2233 (1994); D{\bf 49}, 3352 (1994);
D{\bf 50}, 2225 (1994); K.J. Eskola,
K. Kajantie, P.V. Ruuskanen and K. Tuominen,
Nucl. Phys. {\bf B570}, 379 (2000).

\bibitem{wong}
S. K. Wong, Nuovo Cimento A{\bf 65}, 689 (1970).

\bibitem{heinz}
H.-Th. Elze and U. Heinz, Phys. Rep. {\bf 183}, 81 (1989);
U. Heinz, Ann. Phys. {\bf 161}, 48 (1985); {\bf 168}, 148 (1986).

\bibitem{elze86}
H.-Th. Elze, M. Gyulassy and D. Vasak, Nucl. Phys.
{\bf B276}, 706 (1986).

\bibitem{elze88}
H.-Th. Elze, Z. Phys. C{\bf 38}, 211 (1988). 

\bibitem{mrow} 
S. Mrowczynski, Phys. Rev. D{\bf 39}, 1940 (1989). 


\bibitem{geiger} K. Geiger, Phys. Rev. D{\bf 56}, 2665 (1997).



\bibitem{mgxw} M. Gyulassy and X.-N. Wang,
Nucl. Phys. {\bf B420}, 583 (1994); M. Gyulassy, X.-N. Wang and M.
Pl\"umer, Phys. Rev. D{\bf 51}, 3436 (1995); X.-F. Guo, and X.-N.
Wang, Phys. Rev. Lett. {\bf 85}, 3591 (2000).
%X.-N. Wang, X.-F. Guo, hep-ph/0102230.


\bibitem{baier1} R. Baier et al.,
Nucl. Phys. {\bf B483}, 291 (1997); {\bf B484}, 265 (1997); R.
Baier et al., Nucl. Phys. {\bf B531}, 403 (1997).






\bibitem{dewitt}
B.S. DeWitt, Phys. Rev. D{\bf 162}, 1195 and 1239 (1967).
%; {\it
%in} Dynamic theory of groups and fields (Gordon and Breach, 1965).

\bibitem{thooft}
G. 't Hooft, Nucl. Phys. {\bf B62}, 444 (1973).


\bibitem{abb81} L.F. Abbott, Nucl. Phys. {\bf B185}, 189 (1981);
R.B. Sohn, Nucl. Phys. {\bf B273}, 468 (1986);
H. Kluberg-Stern and J.B. Zuber,
Phys. Rev. D{\bf 12}, 482 (1975).

\bibitem{elze90} H.-Th. Elze, Z. Phys. C{\bf 47}, 647 (1990).

\bibitem{blaizot} J.-P. Blaizot, and E. Iancu,
Nucl. Phys. {\bf B557}, 183 (1999).

\bibitem{blaizot1}
J.-P. Blaizot and  E. Iancu, Nucl. Phys. {\bf B417}, 608 (1994);
{\bf B570}, 326 (2000); Phys. Rev. Lett. {\bf 70}, 3376 (1993).


\bibitem{ctp} J. Schwinger, J. Math. Phys. {\bf 2}, 407 (1961);
L.V. Keldysh, Sov. Phys. JETP {\bf 20}, 1018 (1965); G.Z. Zhou,
Z.B. Su, B.L. Hao and L. Yu, Phys. Rep. {\bf 118}, 1 (1985);
E. Calzetta and B.L. Hu, Phys. Rev. D{\bf 37}, 2878 (1988).

\bibitem{qwang} Q. Wang, K. Redlich, H. St\"ocker and W. Greiner,
in preparation.


\bibitem{pinch} T. Altherr and D. Seibert,
Phys. Lett. B{\bf 333}, 149 (1994);
R. Baier, M. Dirks, K. Redlich and D. Schiff,
Phys. Rev. D{\bf 56}, 2548 (1977);
C. Greiner and S. Leupold, Eur. Phys. J. C{\bf 8}, 517 (1999);
I. Dadic, Phys. Rev. D{\bf 59}, 125012 (1999); D{\bf 63},
025011 (2001).


\end{thebibliography}
\end{document}